 \documentclass[final,3p,times]{elsarticle}

 \usepackage{graphics}
 \usepackage{epstopdf}
 \usepackage{amsmath}
\usepackage{amssymb}
\usepackage{cases}
\usepackage{epsfig,subfigure}
\usepackage[colorlinks=true,linkcolor=blue,citecolor=blue,urlcolor=blue]{hyperref}

 \biboptions{comma,square,compress,numbers,sort}

\journal{Physics Letters B}

\begin{document}

\begin{frontmatter}

\title{ Pure geometric $f(R)$ branes}

\author[a]{Heng Guo\footnote{hguo@xidian.edu.cn, corresponding author}}
\author[a]{Cai-Ling Wang\footnote{wcl@stu.xidian.edu.cn}}
\author[a]{Yong-Tao Lu\footnote{luyt@stu.xidian.edu.cn}}
\author[a]{Yue Sun\footnote{syue@stu.xidian.edu.cn}}
\author[a,b]{Lang-Lang Wang\footnote{wllsrr@163.com}}

\address[a]{School of Physics, Xidian University, Xi'an 710071, China}
\address[b]{School of General Education, Shanxi Institute of Science and Technology, Jincheng 048000, China}

\begin{abstract}

In this paper, we investigate pure geometric $f(R)$ cosmology branes  embedded in five-dimensional spacetime. The form of $f(R)$ is chosen as a polynomial. The Five-dimensional scalar curvature $R$ is assumed to be constant.  Based on the value of the four-dimensional cosmological constant $\lambda_4$, the branes can be classified into Minkowski, de Sitter, and anti-de Sitter cases.
Solutions for each case can be calculated. These solutions are stable against linear tensor perturbations in all cases.
In the Minkowski brane case, the zero mode of gravity can be localized on the brane.
In the de Sitter brane case, the zero mode and one massive Kaluza-Klein mode can be localized on the brane.
In the anti-de Sitter brane case, all massive Kaluza-Klein modes can be localized on the  brane.
The results of the analysis of tensor fluctuations are the same in both the Einstein frame and  the Jordan frame.

  \end{abstract}


\end{frontmatter}


\section{Introduction}

In recent decades, the idea that our observable four-dimensional (4D) universe is a 3-brane embedded in a higher-dimensional spacetime has emerged as a promising framework for addressing longstanding issues in theoretical physics, such as the hierarchy problem and the cosmological constant problem~\cite{VisserPLB1985,AntoniadisPLB1990,LykkenJHEP2000,Maartens2010,AK2002}.
This concept was significantly advanced by Randall and Sundrum~\cite{Randall1999,Randall1999a}, who demonstrated that a massless 4D graviton can emerge on a thin brane provided that a sufficiently large and appropriately warped extra dimension exists. In this Randall-Sundrum (RS) model, observable matter fields are confined to the brane, while gravitons can propagate in the bulk.  However, a zero-thickness brane is an idealization, and more realistic scenarios require branes with thickness. Consequently, thick brane models have been proposed and extensively studied ~\cite{DeWolfe2000a,PRD_koyama,CamposPRL2002,AWang2002,V. Dzhunush,ORI2002,Cvetic,AHA2010,YKY2010,YYK2011,Guo_BentBrane,CCsaki2000,Bazeia2009,Dzhunushaliev2010a,DAR2015,GHMBAE2022}.

A key issue in braneworld scenarios is the localization of gravity and the stability of these solutions.  Numerous studies have explored this issue within the framework of general relativity~\cite{CCsaki2000,Bazeia2009,Koley2005,Gogberashvili2006a,Liu2009,Liu2009a,Guerrero2010,Liu2010a,GogberashviliHerrera-AguilarMalagon-Morejon2010}. However, since general relativity is inherently non-renormalizable, it becomes necessary to consider higher-order curvature terms~\cite{Stelle1977}. Among various extended theories of gravity, $f(R)$ gravity has garnered significant interest for its capability to generalize higher-order gravity while maintaining simplicity and avoiding the Ostrogradski instability~\cite{Woodard2007}.
Originally formulated to address cosmological issues, $f(R)$ gravity offers geometric explanations for phenomena like inflation and dark energy~\cite{VLD1987,JS1988,GESSLS2008,AS2010,TV2010,NDepjc2018,NDplb2018,MYNS2020}.
Numerous researchers have investigated the embedding of branes in various forms of $f(R)$  gravity~\cite{VIAD2007,YYZH2011,YYK2011,DBaRMA2013,DBLLG2014,NA2018,NA2019,AMLZ2016,DBLLRM2015,HYBY2016,MDAVC2015,DDDS2018,SLee2019,AGH2016,KSVW2015,WBFH2018,ACLBj2019,RCAG2019,VDVF2019,YY2016,Cotsakis2023,JDB2020,Chen2021,ZQC2020,BMJN2021,PVCAR2023,GW2023,ACTurrion2023}.
Specifically, Zhong and Liu~\cite{YY2016} obtained solutions for triangular and polynomial $f(R)$ in the context of pure geometric thick $f(R)$ brane scenarios.  The investigation of thick $f(R)$ branes with constant curvature was conducted in Refs.~\cite{VIAD2007,YYK2011}.
The stability of $f(R)$ branes against  linear tensor perturbations was initially addressed in Ref.~\cite{YYK2011} and further explored in subsequent studies~\cite{YY2016,CuiJHEP,ZQC2020,GW2023}. In Ref.~\cite{GW2023}, the authors investigated solutions in pure geometric thick $f(R)$ Minkowski brane models. The solutions are stable against linear tensor perturbations, and the zero modes of gravity can be localized on thick $f(R)$ branes.

Despite extensive research on $f(R)$ gravity, most investigations have only considered the case where the effective 4D metric is Minkowski~\cite{VIAD2007,YY2016,YYK2011,YYZH2011,DBaRMA2013,HYBY2016,ZQC2020,Chen2021,GW2023,ACTurrion2023}, with less attention given to the cosmology of these models.
This paper aims to explore aspects of the cosmology of $f(R)$ branes.
The situation becomes intricate when extending the effective 4D metric from the Minkowski metric to a general metric framework. Therefore, our work starts with the case of five-dimensional (5D) constant curvature, focusing on pure geometric $f(R)$ gravity. More complex scenarios, such as  thick brane solutions with nonconstant curvature in $f(R)$ gravity or $f(R)$ gravity with background scalar fields, are reserved for future research.

This paper is organized as follows.
In Sec.~\ref{section2}, we write down the Einstein equations for a 5D metric based on a general 4D metric, without introducing background scalar fields. By considering $f(R)$ as a polynomial function of the scalar curvature $R$ and constraining $R$ to be constant, we derive  solutions to the equations of motion and discuss their properties in the cases where the 4D brane is Minkowski ($\mathcal{M}_{4}$), de Sitter (dS$_{4}$) and anti-de Sitter (AdS$_{4}$).
In Sec.~\ref{section3}, we discuss tensor perturbations and the localization of gravity. For the $\mathcal{M}_{4}$ brane,  the effective potential closely resembles the potential obtained in the RS model. This  potential involves a $\delta$-function, resulting in a single normalizable bound state mode,  while the remaining eigenstates are continuum modes. For the dS$_4$ brane, there are two bound states. The massless mode represents a localized 4D graviton, while the continuum spectrum begins at $m^2 = \frac{9}{4} H^2$, with modes asymptotically behaving as plane waves representing delocalized KK gravitons. In the case of the AdS$_4$ brane,  the zero mode cannot be localized on the brane, and the ground state is a massive mode with $m_1^{2}=4H^{2}$. All massive modes are bound states localized on the thick AdS$_{4}$ brane, and the  mass spectrum is discontinuous.
In Sec.~\ref{section4}, we transform $f(R)$ gravity into its Einstein frame via conformal transformation to analyze tensor fluctuations,  confirming the  consistency of tensor perturbations between the Einstein and Jordan frames.
Our summary is given in Sec.~\ref{section5}.

\section{The model }\label{section2}

We consider pure $f(R)$ gravity in 5D spacetime as follows:
\begin{eqnarray}
  S = \frac{1}{2\kappa_5^2}\int d^5x\sqrt {-g}f(R),
  \label{action}
\end{eqnarray}
where $R$ is the five-dimensional scalar curvature, $\kappa_5^2 = 8\pi G_5$ with $G_5$ being the 5D Newton constant, and $g = \det(g_{MN})$ is the determinant of the metric. The function $f(R)$ depends on $R$. From this action, the 5D Einstein equation is given by
\begin{eqnarray}
    R_{MN}f_{R} - \frac{1}{2}g_{MN}f(R) + (g_{MN}\Box - \nabla_{M}\nabla_{N})f_{R} = 0, \label{Einstein Eq}
\end{eqnarray}
where $f_{R}\equiv \frac{df(R)}{dR}$, $\Box = g^{MN}\nabla_{M}\nabla_{N}$ is the 5D d'Alembert operator, and $R_{MN}$ is the 5D Ricci tensor, defined in terms of the Riemann tensor as $R_{MN} = R^{Q}_{MQN}$. Throughout this paper, capital Latin letters $M,N,\cdots = 0,1,2,3,5$ and Greek letters $\mu,\nu,\cdots = 0,1,2,3$ are  employed to represent the bulk and brane indices, respectively.

We investigate a warped and static brane, characterized by a metric of the following form:
\begin{eqnarray}
  ds^2 = e^{2A(z)}\left(\hat{g}_{\mu\nu}dx^\mu dx^\nu + dz^2\right).\label{metric}
\end{eqnarray}
Here, $e^{2A(z)}$ represents the warp factor, $\hat{g}_{\mu\nu}$ denotes the induced metric of the 3-brane, and $z$ signifies the extra dimension. For the system described by Eqs. (\ref{action}) to (\ref{metric}), the Ricci curvature tensor and the scalar curvature $R$ can be expressed  as
\begin{eqnarray}
R_{\mu\nu} = \hat{R}_{\mu\nu} - \left(A'' + 3A'^2\right)\hat{g}_{\mu\nu}, \quad\quad
 R_{55} = -4A'' , \label{Ricci tensor}
 \\
\label{scalarR}
R = 4e^{-2A}\left(\lambda_{4}-3A'^2 - 2A''\right).
\end{eqnarray}
Here, the prime denotes derivatives with respect to the extra-dimensional coordinate $z$, and $\hat{R}_{\mu\nu} = \lambda_{4}\hat{g}_{\mu\nu}$, where $\lambda_{4}$ represents the effective 4D cosmological constant obtained after integrating over the fifth dimension~\cite{brane_book}. Depending on the value of $\lambda_{4}$, we can classify the 4D brane into three cases: $\lambda_{4} = 0$ corresponds to a Minkowski brane,  while $\lambda_{4} = 3H^{2}$ and $\lambda_{4} = -3H^{2}$ correspond to dS$_{4}$ and  AdS$_{4}$ branes, respectively. Here, $H$ denotes the Hubble parameter.

By substituting (\ref{Ricci tensor}) and (\ref{scalarR}) into Eq.  (\ref{Einstein Eq}), the Einstein equation can be reformulated as
\begin{subequations}\label{Eeq}
\begin{eqnarray}
   e^{2A}f(R) + 2\left(A'' + 3A'^2 - \lambda_{4}\right)f_R - 4A'f^{\;'}_{R} - 2f^{\;''}_{R} = 0, \label{Einstein Eq1}
  \\
     e^{2A}f(R) + 8A''f_R - 8A'f^{\;'}_{R} = 0. \label{Einstein Eq2}
\end{eqnarray}
\end{subequations}
By organizing and merging the above two equations, we yield
$2\left(3A'^2 - 3A'' - \lambda_{4}\right)f_R + 4A'f^{\;'}_{R} - 2f^{\;''}_{R} = 0$.  In this paper, we consider a 5D constant curvature scenario, which leads to $f^{\;'}_{R} = 0$ and $f^{\;''}_{R} = 0$. Ultimately, we obtain the following equation
\begin{eqnarray}
\left(3A'^2 - 3A'' - \lambda_{4}\right)f_R = 0. \label{Eeq1}
\end{eqnarray}

We aim to investigate the general form of $f(R)$,  which is  chosen as an $n$-th order polynomial of the scalar curvature $R$
\begin{eqnarray}
 f(R) = \sum_{i = 1}^{n} a_{i}\;R^{\,i} - 2 \Lambda_{5}, \label{fR}
\end{eqnarray}
where the coefficients $a_{i}$ possess the appropriate dimensions, $\Lambda_{5}$ denotes the 5D cosmological constant, and the parameter $n$ must be an integer.

\begin{figure}
\begin{center}
\subfigure[]{\label{fig_0Aa}
\includegraphics[width=0.4\textwidth]{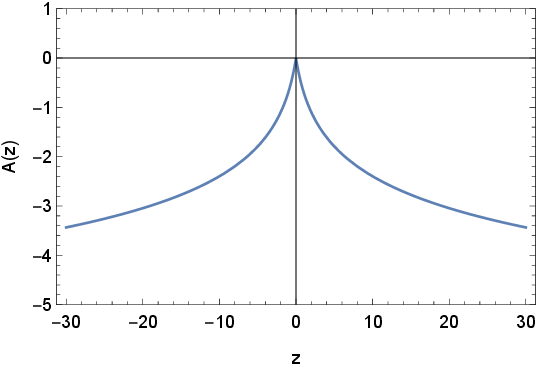}}
\subfigure[]{\label{fig_0e2A}
\includegraphics[width=0.4\textwidth]{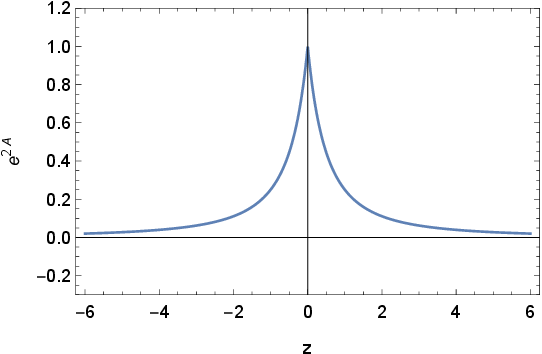}}
 \caption{The shapes of  $A(z)$  in (a) and the warp factor $e^{2A}$ in (b) for the $\mathcal{M}_{4}$ brane .  Here, both $\gamma$ and $k$ are set to 1. }\label{fig_0Ae2A}
\end{center}\vskip -5mm
\end{figure}

In Eq. (\ref{Eeq1}), if $f_R \neq 0$ and $\lambda_4 = 0$, this indicates a $\mathcal{M}_{4}$ brane with zero curvature. The following expressions can be obtained
\begin{eqnarray}
         A(z) = -\ln (\gamma + k |z|),  \quad R = -20 k^{2} + 16k(\gamma +  k |z|)\delta(z), \label{R000}
\end{eqnarray}
where $\gamma$ is a dimensionless constant and $k$ is another constant with the dimension of length inverse. The graphical representations of $A(z)$ and the corresponding warp factor $e^{2A(z)}$ are presented  in Figure~\ref{fig_0Ae2A}. It is noteworthy that the function $A(z)$ in this scenario is not smooth; it contains a cusp at $z=0$. This cusp in $A(z)$ results in the warp factor $e^{2A(z)}$ being discontinuous and exhibiting at most a $\delta$-function in the the context of second-order Einstein gravity, which can be explained as the emergence of a thin brane. The expression of the 5D scalar curvature $R$ incorporates a $\delta$-function. By introducing the tension of the brane, the RS braneworld model ~\cite{Randall1999,Randall1999a} can be concluded.

\begin{figure}
\begin{center}
\subfigure[]{\label{fig_dSA}
\includegraphics[width=0.4\textwidth]{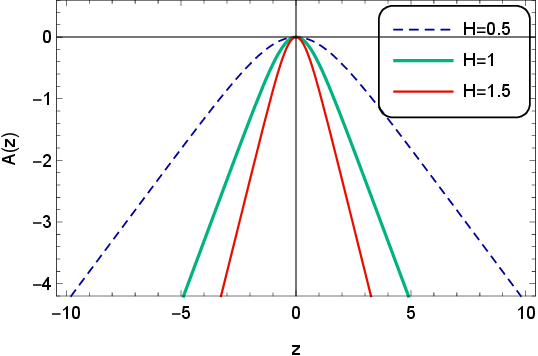}}
\subfigure[]{\label{fig_dSe2A}
\includegraphics[width=0.4\textwidth]{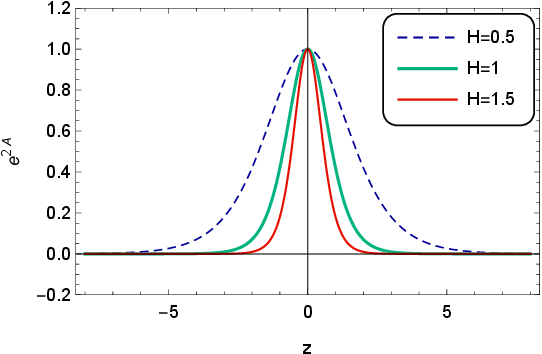}}
 \caption{ The shapes of  $A(z)$  in (a) and the warp factor $e^{2A}$ in (b) for the  $\mathrm{dS_{4}}$ brane vary with different values of $H$.  The dashed blue line represents $H = 0.5$, the green line represents $H = 1$, and  the thin red line represents $H= 1.5$. }\label{fig_dSAe2A}
\end{center}\vskip -5mm
\end{figure}

In Eq. (\ref{Eeq1}), if $f_R \neq 0$, $\lambda_{4} = 3H^{2}$ and $\lambda_{4} = -3H^{2}$ correspond to dS$_{4}$ and AdS$_{4}$ branes, respectively.
For the dS$_{4}$ brane, the following expressions can be obtained
\begin{align}
         A(z)& = \ln (\text{sech} (Hz)),\quad R = 20 H^{2}, \label{R10}
\\
         a_{i}& = 2\alpha\frac{1}{i!}\left( \frac{1}{8H^{2}} \right)^{i}, \\
  \Lambda_{5}& = \alpha\left[\frac{1}{n!}\left(\frac{5}{2}\right)^{n} - 1\right], \label{13}
\end{align}
where $\alpha$ is a dimensionless constant. Note that in this scenario, the function $A(z)$ is negative across all values of $z$, except at $z = 0$, where it approaches negative infinity as $z$ tends towards $\pm\infty$.  The graphical representations of $A(z)$ and the corresponding warp factor $e^{2A(z)}$ are depicted  in Figure~\ref{fig_dSAe2A}. The 5D scalar curvature $R$ is positive.  We assume $\alpha > 0$ to ensure $a_1 > 0$. Consequently, for $0 \leqslant n \leqslant 4$, the cosmological constant satisfies $\Lambda_{5} > 0$, indicating that the 5D spacetime corresponds to de Sitter (dS$_{5}$) spacetime. In contrast, for $n \geqslant 5$, $\Lambda_{5} < 0$ implying that the 5D spacetime is characterized as anti-de Sitter (AdS$_{5}$) spacetime.

When $n = 1$ and $\alpha = 4 H^2$, $f(R) = R - 2 \Lambda_{5}$, where $\Lambda_{5} = 6 H^2$. From this, we obtain $S = \frac{1}{2\kappa_5^2}\int d^5x\sqrt {-g}(R - 12 H^2)$. The corresponding Einstein equation with a 5D cosmological constant is $R_{MN} - \frac{1}{2}g_{MN}R = -\Lambda_{5}g_{MN}$. It is evident that, in this case, $f(R)$ gravity reduces to gravity within the framework of general relativity. This corresponds to the case when $b=H$  in  Ref.~\cite{HGUO2013}.  In this scenario,  both cosmological constants  $\lambda_{4}$ and $\Lambda_{5}$ are positive, which accounts for
the embedding of a dS$_{4}$ brane into a chart of the dS$_{5}$ spacetime.

\begin{figure}
\begin{center}
\subfigure[]{\label{fig_AdSA}
\includegraphics[width=0.4\textwidth]{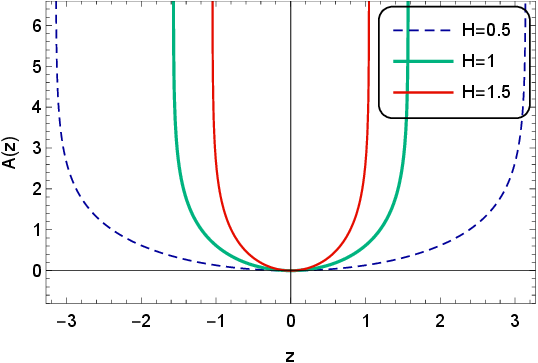}}
\subfigure[]{\label{fig_AdSe2A}
\includegraphics[width=0.4\textwidth]{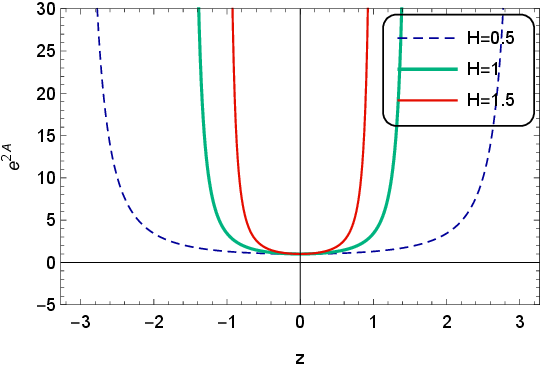}}
 \caption{The shapes of $A(z)$  in (a) and the warp factor $e^{2A}$ in (b) for the $\mathrm{AdS_{4}}$ brane vary with different values of $H$. The dashed blue line represents $H = 0.5$, the green line represents $H = 1$, and  the thin red line represents $H = 1.5$.}\label{fig_AdSAe2A}
\end{center}\vskip -5mm
\end{figure}

In the case of the AdS$_{4}$ brane, we obtain the following result:
\begin{align}
         A(z)& = \ln (\sec (Hz)),\quad
         R = -20 H^{2},\label{R13}
\\
         a_{i}& = 2\beta\frac{1}{i!}\left(- \frac{1}{8H^{2}} \right)^{i}, \\ \Lambda_{5}& = \beta\left[ \frac{1}{n!}\left(\frac{5}{2}\right)^{n} - 1\right
], \label{17}
\end{align}
where $\beta$ is a dimensionless constant. Here, we consider the range for $z$ to be $- \frac{\pi}{2H} <  z  < \frac{\pi}{2H}$. Notably, in the case of the AdS$_{4}$ brane,  the function $A(z)$ is positive except at $z = 0$, where it approaches positive infinity as $z$ tends towards $\pm\frac{\pi}{2H}$.  The graphical representations of $A(z)$ and the corresponding warp factor $e^{2A(z)}$ are depicted  in Figure~\ref{fig_AdSAe2A}. The  5D scalar curvature $R$ is negative.  We assume $\beta < 0$ to ensure $a_1 > 0$. Therefore,  when $0 \leqslant  n \leqslant 4 $,  the cosmological constant $\Lambda_{5} < 0$,  indicating that the 5D spacetime is AdS$_{5}$; conversely, when $n \geqslant 5$, $\Lambda_{5} > 0$, suggesting that the
 5D spacetime corresponds to dS$_{5}$.

When $n = 1$ and $\beta = -4 H^2$, $f(R) = R - 2\Lambda_{5}$, where $\Lambda_{5} = -6 H^2$. We obtain $S = \frac{1}{2\kappa_5^2}\int d^5x\sqrt {-g}(R + 12 H^2)$, and the Einstein equation with a 5D cosmological constant is $R_{MN} - \frac{1}{2}g_{MN}R = 6H^{2}g_{MN}$.

In Table \ref{table ai}, we present the values of $a_{1} \thicksim a_{6}$ and $a_{10}$ associated with the dS$_{4}$ brane and AdS$_{4}$ brane. Additionally, we include the values of $\Lambda_{5}$ for $n$ ranging from $1$ to $6$, and {10}. From Eq. (\ref{13}) and Eq. (\ref{17}), we observe that the expressions for the 5D cosmological constant $\Lambda_{5}$ have the same form in both cases, and as  $n \rightarrow {\infty}$, $\Lambda_{5}$ approaches -$\alpha$ and -$\beta$ respectively. Notably, we find that when $n \geqslant 5$, the sign of the 5D cosmological constant $\Lambda_{5}$ is opposite to that of the 4D brane cosmological constant, suggesting that the 4D brane cosmological constant is independent of the 5D cosmological constant.

\begin{table}[tbp]
\centering
\renewcommand{\arraystretch}{1.5} 
\begin{tabular}{|c|c|c||c|c|c|}
\hline
    $i$    &$a_{i(\mathrm{dS}_{4})}$    & $a_{i(\mathrm{AdS}_{4})}$   & $n$  &  $\Lambda_{5(\mathrm{dS}_{4})}$  & $\Lambda_{5(\mathrm{AdS}_{4})}$\\
\hline
    $1$ & $\frac{\alpha}{4H^2}$ & $-\frac{\beta}{4H^2}$ & $1$ & $\frac{3}{2}\alpha$ &$\frac{3}{2}\beta$\\
\hline
    $2$  &  $\frac{\alpha}{64H^4}$ &  $\frac{\beta}{64H^4}$ &  $2$ & $\frac{17}{8}\alpha$  & $\frac{17}{8}\beta$\\
\hline
    $3$  &  $\frac{\alpha}{1536H^6}$ &  $-\frac{\beta}{1536H^6}$ &  $3$ & $\frac{77}{48}\alpha$  &  $\frac{77}{48}\beta$\\
\hline
    $4$  & $\frac{\alpha}{49152H^8}$ & $\frac{\beta}{49152H^8}$ &    $4$ & $\frac{241}{384}\alpha$ &$\frac{241}{384}\beta$  \\
\hline
    $5$  & $\frac{\alpha}{1966080H^{10}} $   &  $-\frac{\beta}{1966080H^{10}} $ & $5$ &   $-\frac{259}{384}\alpha$  & $-\frac{259}{384}\beta$ \\
\hline
    $6$  & $\frac{\alpha}{94371840H^{12}}$& $\frac{\beta}{94371840H^{12}}$               &  $6$ & $-\frac{6091}{9216}\alpha$ &  $-\frac{6091}{9216}\beta$\\
\hline
    $10$  &  $\frac{\alpha}{1948197165465600H^{20}}$  &   $\frac{\beta}{1948197165465600H^{20}}$             &  $10$ &   $-\frac{148245023}{148635648}\alpha$&  $-\frac{148245023}{148635648}\beta$\\
\hline
\end{tabular}
\caption{\label{table ai} The values of $\mathit{a_{i}}$ and $\mathrm{\Lambda_{5}}$ corresponding to the $\mathrm{dS_{4}}$ and $\mathrm{AdS_{4}}$ branes.}
\end{table}

\section{Tensor perturbations and localization of gravity  }\label{section3}

Within the braneworld framework, the localization of gravity and bulk matter fields is crucial for reproducing effective 4D gravity and formulating the standard model. The localization of zero modes reproduces effective 4D gravity, while massive KK modes modify Newtonian gravity on the brane~\cite{CCsaki2000,Giovannini2001,Bozza2001}.
Furthermore, stability under small fluctuations of all fields is a fundamental requirement for any consistent braneworld model~\cite{PRD_koyama}.

In the preceding section, three distinct types of
brane systems were constructed. We now shift our focus to small metric fluctuations around the solutions derived in the previous section. Our objective is to demonstrate that all these solutions are stable against metric perturbations to linear order.
We consider the following metric perturbations:
\begin{eqnarray}
ds^2 = e^{2A(z)}\left[\left(\hat{g}_{\mu\nu} + h_{\mu\nu}\right)dx^\mu dx^\nu + dz^2 \right]. \label{ds20}
\end{eqnarray}
Alternatively, in another form
\begin{eqnarray}
g_{MN} = \bar{g}_{MN} + \Delta g_{MN}
 = e^{2A(z)} \left(
  \begin{array}{cc}
    \hat{g}_{\mu\nu} & 0 \\
    0 & 1 \\
  \end{array}
\right) + e^{2A(z)}\left(
  \begin{array}{cc}
   h_{\mu\nu} & 0 \\
    0 & 0 \\
  \end{array}
\right),
\end{eqnarray}
where $h_{\mu\nu} = h_{\mu\nu}(x^{\rho},~z)$ depends on all the
coordinates. Clearly, $\Delta g_{5M} = 0$, indicating that we consider
only tensor perturbations. From the relation
$g^{NP}g_{PM} = \delta_M^{~N}$, we obtain the inverse of $\Delta
g_{MN}$, denoted as $\Delta g^{MN}$. We retain only the first-order term,
i.e., $\Delta g^{MN(1)}$, denoted as
\begin{eqnarray}
\delta g^{MN} = e^{-2A(z)}\left(
  \begin{array}{cc}
    -h^{\mu\nu} & 0 \\
    0 & 0 \\
  \end{array}
\right),
\end{eqnarray} where $h^{\mu\nu} = \hat{g}^{\mu\lambda}\hat{g}^{\nu\rho}h_{\lambda\rho}$ is
raised using $\hat{g}^{\mu\nu}$. We consistently use $\delta X$ to denote the first-order contribution of perturbations
to any arbitrary quantity $X$.
We can then derive the perturbations of the Ricci curvature tensor and the scalar curvature
\begin{align}
 \delta R_{\mu\nu}& = \frac{1}{2}\left[\hat{\nabla} _{\mu }\hat{\nabla} _{\sigma}h_{\nu }^{\sigma } + \hat{\nabla} _{\nu }\hat{\nabla} _{\sigma}h_{\mu }^{\sigma } - \hat{\square}^{(4)} h_{\mu \nu } - \hat{\nabla} _{\mu }\hat{\nabla} _{\nu }h\right]
                - \frac{3}{2} A' h_{\mu \nu }' - \frac{1}{2} h_{\mu \nu }''\nonumber\\
               &\quad - \frac{1}{2} A'\hat{g}_{\mu\nu}h' + \frac{4}{3}\lambda_{4}h_{\mu \nu }-\frac{\lambda_{4}}{3}\hat{g}_{\mu\nu}h -  A''h_{\mu \nu } - 3A'^{2}h_{\mu \nu } ,\nonumber\\
\delta R_{\mu 5}& = \frac12\left(\hat{\nabla} _{\alpha}
h'^{\,\alpha}_{\mu} - \hat{\nabla} _{\mu}
h'\right),\quad
  \delta R_{55}=-\frac{1}{2}A' h' - \frac{1}{2}h'' ,\nonumber\\
    \delta R &= e^{-2A}\left(\hat{\nabla} _{\mu }\hat{\nabla} _{\nu}h^{\mu\nu} - \hat{\square}^{(4)} h - 4A' h' - A''h - h'' - 2A''h - 6A'^{2}h + \lambda_{4}h\right).
\label{relationsR}
\end{align}
Here $\hat{\square}^{(4)} = \hat g^{\mu\nu}\hat{\nabla} _{\mu }\hat{\nabla} _{\nu}$ represents
the 4D d'Alembert operator, while $h = \hat g^{\mu\nu}h_{\mu
\nu}$ denotes the trace of the tensor perturbations. By applying the transverse-traceless (TT) condition $h = 0 = \hat{\nabla} _{\mu } h^\mu_{~\nu}$, only $\delta R_{\mu\nu}$ remains non-zero, significantly simplifying the perturbed equations.

In $f(R)$ gravity, when considering the perturbations, the feedback from the Einstein equations, as shown in Eq. (\ref{Einstein Eq}), is expressed as
\begin{eqnarray}
 [(\delta R_{MN}) - \frac{1}{2} g_{MN}(\delta R) + (\delta g_{MN})    \square^{(5)} + g_{MN}(\delta g^{AB})\nabla_{A} \nabla_{B} - g_{MN}g^{AB}(\delta\Gamma^P_{AB})\nabla_{P}  \nonumber\\
 + (\delta\Gamma^P_{MN})\nabla_{P}]f_{R} - \frac{1}{2}(\delta g_{MN})f(R) + [ R_{MN}(\delta R) + g_{MN}\square^{(5)}(\delta R) + g_{MN}(\delta R)\square^{(5)}  \nonumber\\
 + 2g_{MN}g^{AB}(\nabla_{A}\delta R)\nabla_{B} - \nabla_{M} \nabla_{N}\delta R - (\delta R)\nabla_{M} \nabla_{N} - (\nabla_{N}\delta R)\nabla_{M} - (\nabla_{M}\delta R)\nabla_{N}] f_{RR} = 0.
 \label{eqPertubedEE}
\end{eqnarray}
By substituting Eq. (\ref{relationsR})  into Eq. (\ref{eqPertubedEE})  and applying the TT condition, we derive the perturbed Einstein equations as follows
\begin{eqnarray}
(\delta R_{\mu5})f_{R} - \left[\hat{\nabla} _{\mu}(\delta R) - A'\hat{\nabla} _{\mu}(\delta R)\right]f_{RR}-\hat{\nabla} _{\mu}(\delta R)f'_{RR} = 0,\label{ee2}
\\
 \left(\delta R_{55} - \frac{1}{2}e^{2A}\delta R\right)f_{R} + \left[R_{55}\delta R + \square^{(4)}(\delta R) + 4A'(\delta R)'\right]f_{RR}
 + \left(\frac{1}{2}h' + 2A'h\right)f'_{R} + 4A'(\delta R)f'_{RR} = 0, \label{ee3}
\\
\left(-\frac{1}{2}\hat{\square}^{(4)}h_{\mu\nu} - \frac{3}{2}A'h'_{\mu\nu}- \frac{1}{2}h''_{\mu\nu} + \frac{1}{3}\lambda_{4}h_{\mu\nu} \right)f_{R} - \frac{1}{2}h'_{\mu\nu}f'_{R} = 0.
\label{relationsRuv}
\end{eqnarray}

To demonstrate the occurrence of gravitational localization, we examine the zero mode, along with its normalization and asymptotic behavior. Additionally, we analyze the asymptotic behavior of the effective potential. Therefore, we express Eq. (\ref{relationsRuv})  in the following form
\begin{eqnarray}
 \left[\partial_z^{~2}
  + \left(3A' + \frac{f'_R}{f_R}\right)\partial_z
  + \hat{\square}^{(4)} - \frac{2}{3}\lambda_{4}\right]h_{\mu\nu} = 0.
\end{eqnarray}
We consider the decomposition
$h_{\mu\nu}(x^{\rho}, z) = \epsilon_{\mu\nu}(x^{\rho})e^{-\frac{3}{2}A}f_R^{-1/2}\psi(z)$ and require that $\epsilon_{\mu\nu}(x^{\rho})$ satisfies the TT condition $g^{\mu\nu}\epsilon_{\mu\nu} = 0 = \partial_\mu
\epsilon^{~\mu}_\nu$. This leads to the 4D component equation $\left(\hat{\square}^{(4)} - \frac{2}{3}\lambda_{4}\right)\epsilon_{\mu\nu}(x^{\rho}) = m^{2}\epsilon_{\mu\nu}(x^{\rho})$. Subsequently, a Schr\"odinger-like equation for $\psi(z)$ emerges
\begin{eqnarray}
  \left[-\partial_z^2
      + W(z)\right]\psi(z)
      = m^2\psi(z),\label{Schrodinger}
\end{eqnarray}
where we drop the $\mu\nu$ indices from the wavefunctions, which are now labeled by the corresponding energy $m^{2}$. The effective potential $W(z)$ is given by
\begin{eqnarray}
 W(z) = \frac94A'^2
       + \frac32A''
       + \frac32A'\frac{f'_R}{f_R}
       - \frac14\frac{f_R'^2}{f_R^2}
       + \frac12\frac{f_R''}{f_R}.
      \label{Schrodingerpotential}
\end{eqnarray}
By constraining $R$ to be a constant, this equation simplifies to
\begin{eqnarray}
W(z) = \frac94A'^2
       + \frac32A''.\label{ww}
\end{eqnarray}
It is clear that the potential  $W(z)$ is entirely determined by the warp factor;  thus, it represents a 4D mass-independent potential.  Notably,  this takes the same form as the case of graviton KK modes in general relativity~\cite{Bazeia2009}.  As reported in Refs. \cite{CCsaki2000,Brandhuber_JHEP}, to
localize the massless gravity mode on a brane, the potential  $W(z)$ must exhibit a well with a negative minimum within the brane's vicinity and adhere to the condition $W_{z}\geq 0$ as $|z| \rightarrow \infty$. For the distinct solutions of the warp factor $e^{2A}$ corresponding to Minkowski branes, dS$_{4}$ and AdS$_{4}$ branes, the effective potentials exhibit different specific expressions.

For clarity, the Schr\"odinger-like
equation (\ref{Schrodinger}) can also be factorized as
\begin{eqnarray}
QQ^{\dagger}\psi(z) = m^2\psi(z),
\end{eqnarray}
where
\begin{eqnarray}
Q = + \partial _z
  + \left(\frac{3}{2}\partial_zA + \frac{1}{2}\frac{\partial_z f_R}{f_R}\right),\nonumber\\
Q^{\dagger} = -\partial _z + \left(\frac{3}{2}\partial_zA + \frac{1}{2}\frac{\partial_z f_R}{f_R}\right),
\end{eqnarray}
indicating that there
are no gravitational modes with $m^2 < 0$. Consequently, any solution of the
system (\ref{action})-(\ref{metric}) is stable against tensor
perturbations.

The zero mode is expressed as
\begin{eqnarray}
\psi_{(0)}(z) = N_{G}e^{\frac{3}{2}A}f_R^{\frac{1}{2}}, \label{zeromode}
\end{eqnarray}
 where $N_G$ represents the normalization constant. The constant  $N_G$ can be determined from the orthogonal normalization condition
\begin{eqnarray}
1 = \int_{-\infty}^{+\infty}|\psi_{(0)}(z)|^2 dz.\label{occ}
\end{eqnarray}

Subsequently, we will conduct a separate analysis of the effective potentials associated with three distinct types of branes.

\subsubsection*{\textbf{\emph{Case 1: $\mathcal{M}_{4}$ brane}}}

\begin{figure}
\begin{center}
\subfigure[]{\label{fig_flatW}
\includegraphics[width=0.4\textwidth]{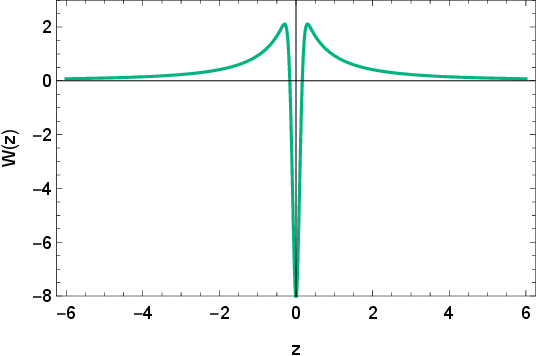}}
\subfigure[]{\label{fig_flatPsi0}
\includegraphics[width=0.4\textwidth]{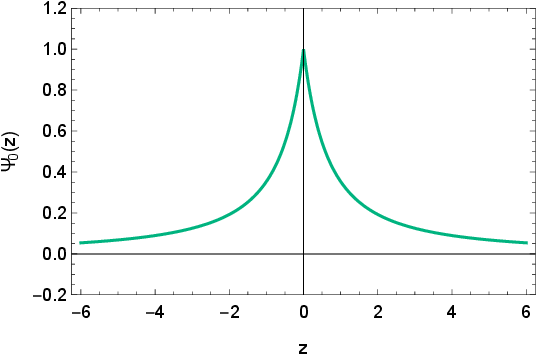}}
 \caption{ The shapes of the potential $W(z)$ and  the zero mode $\psi_{0}(z)$ for the $\mathcal{M}_{4}$ brane. Here, both $\gamma$ and $k$ are set to 1. The chosen form of $\delta(z)$ is given by $\delta(z) = \frac{7}{\sqrt{\pi}}e^{-49 z^2}$.}\label{fig_flatz}
\end{center}\vskip -5mm
\end{figure}

For the $\mathcal{M}_{4}$ brane, Eq. (\ref{ww}) provides the corresponding effective
potential
\begin{eqnarray}
W(z) = \frac{15}{4}\frac{k^2}{(\gamma + k|z|)^2} - \frac{3k}{\gamma + k|z|} \delta(z).\label{w0}
\end{eqnarray}
This effective potential closely resembles the potential obtained in the RS model in Ref.~\cite{Randall1999}. The $\delta$-function supports a single normalizable bound state mode, while the remaining eigenstates are continuum modes. Given the explicit form of the KK potential, we can  gain insights into the properties of the continuum modes. Firstly, since the potential approaches zero as $|z| \rightarrow \infty$, there is no gap, and the continuum modes asymptote to plane waves. Additionally, the amplitudes of the continuum modes are suppressed near the origin due to the potential barrier near $z = 0$. Lastly, the continuum KK states can possess all values of  $m^2 > 0$.  The potential $W(z)$ and  the zero mode $\psi_{0}(z)$ are depicted in Figure~\ref{fig_flatz}.

\subsubsection*{\textbf{\emph{Case 2:  dS$_{4}$ brane}}}

\begin{figure}
\begin{center}
\includegraphics[width=0.4\textwidth]{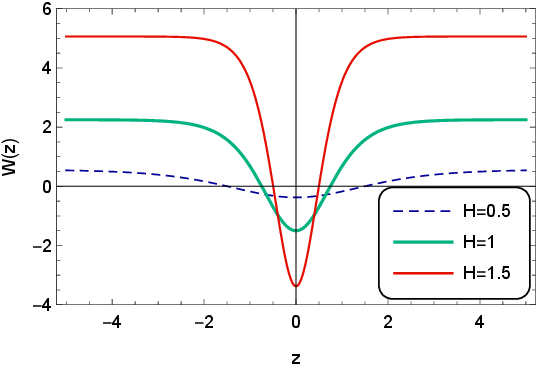}
\caption{ The shapes of the potential $W(z)$ for the $\mathrm{dS_{4}}$ brane vary with different values of $H$, where the dashed blue line represents $H = 0.5$, the green line represents  $H=1$, and the thin red line represents $H = 1.5$.}\label{fig_dSW}
\end{center}
\end{figure}

\begin{figure}
\begin{center}
\subfigure[]{\label{fig_dSPsi0}
\includegraphics[width=0.4\textwidth]{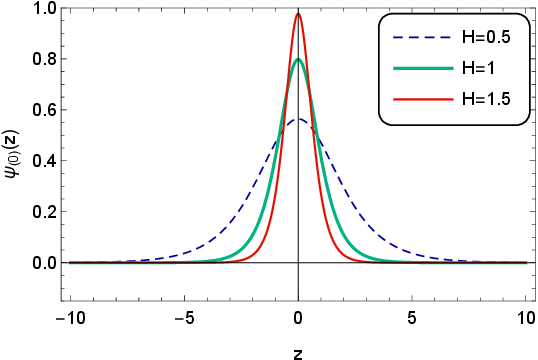}}
\subfigure[]{\label{fig_dSPsi1}
\includegraphics[width=0.4\textwidth]{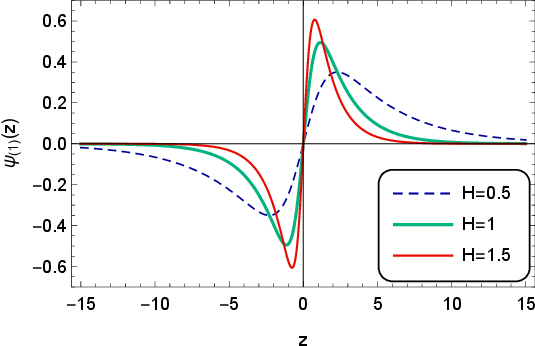}}
 \caption{ The shapes of the zero mode in (a) and the first mass mode in (b) for the $\mathrm{dS_{4}}$ brane vary with different values of $H$, where the dashed blue line represents $H = 0.5$, the green line represents $H = 1$, and the thin red line represents $H = 1.5$.}\label{fig_dSPsi}
\end{center}\vskip -5mm
\end{figure}

For the dS$_{4}$ brane,  the effective potential given by Eq. (\ref{ww}) is expressed as
\begin{eqnarray}
W(z) = \frac94H^2 - \frac{15}{4}H^{2}\textrm{sech}^2(Hz),\label{w1}
\end{eqnarray}
which represents a P\"oschl-Teller potential, as depicted in Figure ~\ref{fig_dSW}. This potential exhibits  a minimum value of $-\frac{3}{2}H^2$ at $z = 0$ and a maximum value of $\frac{9}{4}H^2$ as $z = \pm\infty$, thereby ensuring the existence of a mass gap in the spectrum.  The analytic structure of $W(z)$ facilitates the investigation of gravity localization. Specifically, the normalized zero mode is given by
\begin{eqnarray}
\psi_{(0)}(z) = \sqrt{\frac{2H}{\pi}}\textrm{sech}^{\frac{3}{2}}(Hz),\label{p0}
\end{eqnarray}
as shown in Figure~\ref{fig_dSPsi0}.

\begin{figure}
\begin{center}
\includegraphics[width=0.4\textwidth]{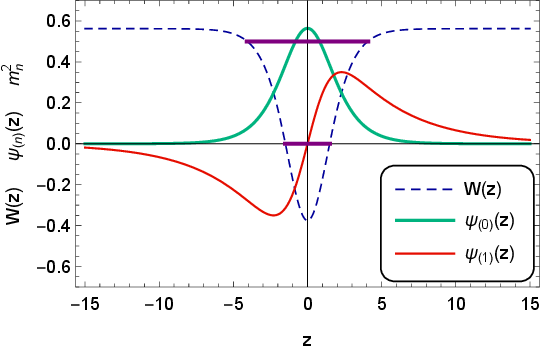}
\caption{The shape of KK modes $W(z)$ (dashed blue line) , the bound KK modes  $\psi_{(n)}(z)$ (green line for $\psi_{(0)}(z)$ and  thin red line for $\psi_{(1)}(z)$), and the mass spectrum (thick purple lines) of the $\mathrm{dS_{4}}$ brane when $H = 0.5$. }\label{fig_dSWPsim}
\end{center}
\end{figure}

Therefore, Eq. (\ref{Schrodinger}) can be expressed as
\begin{eqnarray}
\left[-\partial_z^2
       - \frac{15}{4}H^{2}\textrm{sech}^2(Hz)\right]\psi_{(n)}(z)
       = E_{n}\psi_{(n)}(z),
\end{eqnarray}
where  $E_n = m_{n}^2 - \frac94H^2$. Then, the energy spectrum of the bound state  and  the squared mass can be derived:
\begin{eqnarray}
E_n = -\left(\frac32-n\right)^{2}H^2 ,\quad &&
m_{n}^{2} = n(3 - n)H^2.
\end{eqnarray}
Here, $n$ is an integer that satisfies $0\leqslant n \leqslant {\frac32}$, indicating that  there are two bound states. The first bound state corresponds to $n = 0$, with its wave function expressed  as
\begin{eqnarray}
\psi_{(0)}(z) = \sqrt{\frac{2H}{\pi}}\textrm{sech}^{\frac{3}{2}}(Hz) ,
\end{eqnarray}
which corresponds to Eq.(\ref{p0}) and represents the massless mode  localized as the 4D graviton on the dS$_{4}$ brane. The second bound state is the first excited state, corresponding to $n = 1$ and $m_1^{2} = 2H^2$. Its wave function can be expressed as
\begin{eqnarray}
\psi_{(1)}(z) = \sqrt{\frac{2H}{\pi}}\textrm{sech}^{\frac{3}{2}}(Hz)\sinh(Hz).
\end{eqnarray}
The shapes of the two bound KK modes and the corresponding mass spectrum are depicted in Figure~\ref{fig_dSPsi} and Figure~\ref{fig_dSWPsim}. The continuous spectrum
starts at  $m^2 = \frac{9}{4}H^2 $ and is described by the following eigenfunctions with
imaginary order $\mu = i\rho$ \cite{NAMC2008}
\begin{equation}
\psi_{(n)}(z) = C_{1}\,P^{\pm i\rho}_{\frac{3}{2}}\left(\tanh(Hz)
\right) + C_{2}\,Q^{\pm i\rho}_{\frac{3}{2}}\left(\tanh(Hz) \right).
\label{massmod}
\end{equation}
Here, $P^{\mu}_{\frac{3}{2}}$ and $Q^{\mu}_{\frac{3}{2}}$ are the associated Legendre functions of the first and second kind,
respectively, of degree $\nu = \frac32$, with  $\rho = \sqrt{\frac{n^{2}}{H^{2}} - \frac{9}{4}}$. These states asymptotically turn into plane waves and represent delocalized KK massive gravitons. Thus, when the energy of the scalars is larger than  $\frac{3}{2}H $ , the scalars cannot be trapped on the
brane; instead, they will be excited into the bulk, indicating a potential opportunity for experimentally discovering extra dimensions.

\subsubsection*{\textbf{\emph{Case 3: AdS$_{4}$ brane}}}

\begin{figure}
\begin{center}
\subfigure[]{\label{fig_AdSW}
\includegraphics[width=0.4\textwidth]{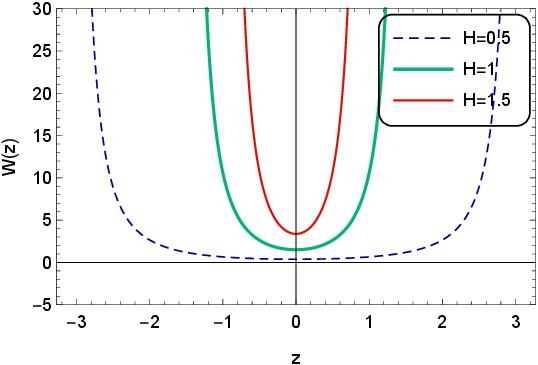}}
\subfigure[]{\label{fig_AdSPsi0}
\includegraphics[width=0.4\textwidth]{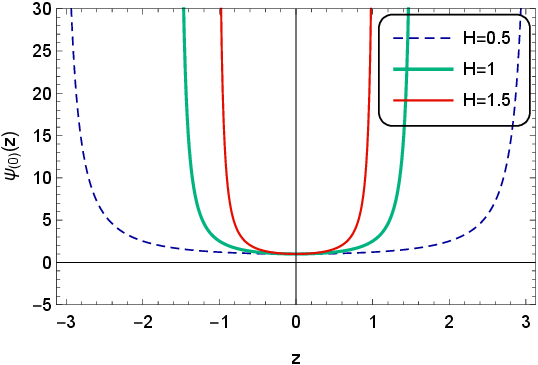}}
 \caption{ The shapes of the potential $W(z)$ in (a) and the non-normalized  zero mode $\psi_{(0)}(z)$ in (b) for the $\mathrm{AdS_{4}}$ brane vary with different values of $H$, where  the dashed blue line represents $H = 0.5$, the green line represents $H = 1$, and  the thin red line represents $H = 1.5$. Here, $N_{G(\text{AdS}) }$ is set to 1.}\label{fig_AdSPsi}
\end{center}\vskip -5mm
\end{figure}

For the AdS$_{4}$ brane, Eq. (\ref{ww})  provides the corresponding Schr\"odinger-like
potential
 \begin{eqnarray}
W(z) = -\frac94H^2
      + \frac{15}{4}H^2\textrm{sec}^2(Hz).\label{-3p}
\end{eqnarray}
This potential function is periodic with a period of $\frac{k\pi}{H}( k\in\text{Z})$. For our analysis, we restrict our focus to the interval $- \frac{\pi}{2H} < z < \frac{\pi}{2H}$ for this potential function and present its graphical representation in Figure~\ref{fig_AdSW}. Notably, the potential function attains its minimum value of $\frac{3}{2}H^2$ at  the brane location ($z = 0$) and approaches positive infinity at the boundaries of the fifth dimension. Consequently, the potential does  not assume negative
values at the brane location, rendering it incapable of trapping the zero mode of gravity, while it can trap the massive modes.

Given that $R$ is a constant, the zero mode can be expressed as
\begin{eqnarray}
\psi_{(0)}(z) = N_{G(\text{AdS}) }\textrm{sec}^{\frac32}(Hz),
\end{eqnarray}
where $N_{G(\text{AdS}) }
$ is a constant; however, the zero mode does not satisfy the orthogonal convergence condition defined in Eq. (\ref{occ}). The non-normalized zero mode in the interval $-\frac{\pi}{2H} < z < \frac{\pi}{2H}$ is shown in Figure~\ref{fig_AdSPsi0} and diverges at the period boundary. Therefore, the zero mode cannot be trapped on the AdS$_{4}$ thick brane.

\begin{figure}
\begin{center}
\subfigure[$n=1$]{\label{fig_AdSPsi1}
\includegraphics[width=0.23\textwidth]{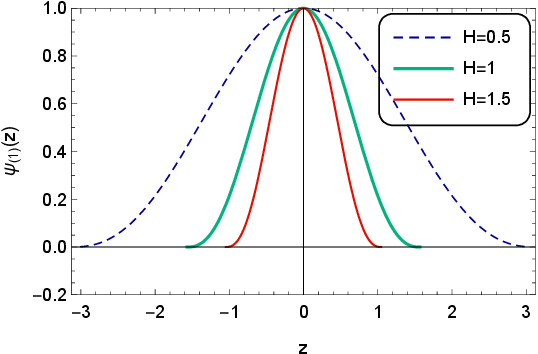}}
\subfigure[$n=2$]{\label{fig_AdSPsi2}
\includegraphics[width=0.23\textwidth]{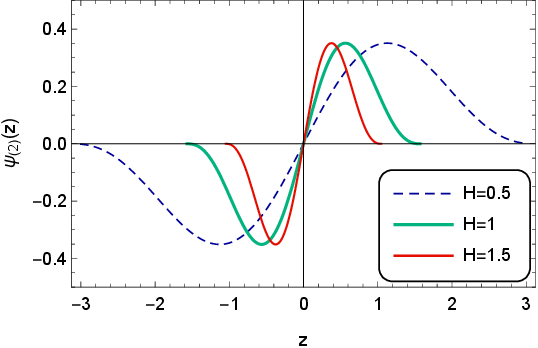}}
\subfigure[$n=3$]{\label{fig_AdSPsi3}
\includegraphics[width=0.23\textwidth]{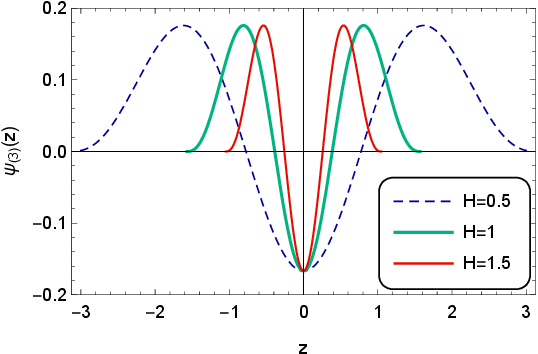}}
\subfigure[$n=4$]{\label{fig_AdSPsi4}
\includegraphics[width=0.23\textwidth]{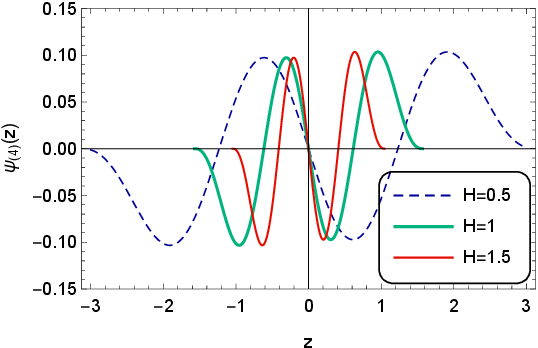}}
 \caption{ The shapes of the bound KK modes $\psi_{(n)}(z)$ for the $\mathrm{AdS_{4}}$ brane  vary with different values of $H$, where  the dashed blue line represents $H = 0.5$, the green line represents $H = 1$, and the thin red line represents $H = 1.5$.   Here, $C$ is set to 1.}\label{fig_AdSPsin}
\end{center}\vskip -5mm
\end{figure}

However, based on the potential  $ W(z)$ in Eq. (\ref{-3p}), there exist an infinite number of bound KK modes. In this context,  Eq. (\ref{Schrodinger}) can be expressed as
\begin{eqnarray}
\left[-\partial_z^2
     + \frac{15}{4}H^{2}\textrm{sec}^2(Hz)\right]\psi_{(n)}(z)
       = E_{n}\psi_{(n)}(z),
\end{eqnarray}
where  $E_{n} = m_{n}^2 + \frac94H^2 $. The solution for the gravity KK modes is
\begin{eqnarray}
\psi_{(n)}(z) = C \cdot {}_2F_1\left(1 - n, 4 + n; 3; \frac{1 - \sin(Hz)}{2}\right)\cdot\cos^{\frac52}(Hz),
\end{eqnarray}
where $C$ is a constant, and $n = 1, 2 \cdots$. The shape of the lower bound KK modes is shown in Figure ~\ref{fig_AdSPsin}.

\begin{figure}
\begin{center}
\subfigure[$H=0.5$]{\label{fig_AdSmH05}
\includegraphics[width=0.23\textwidth]{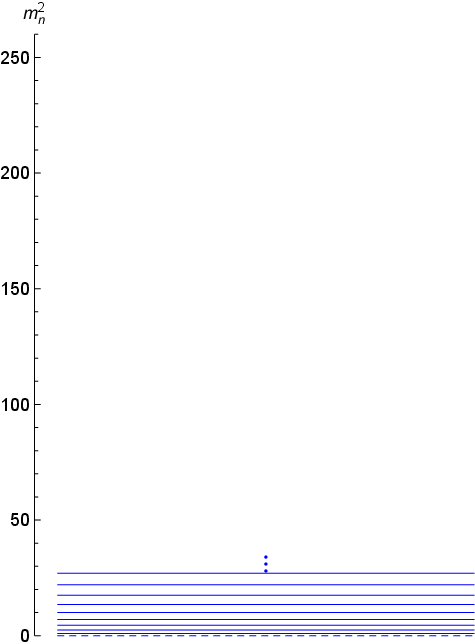}}
\subfigure[$H=1$]{\label{fig_AdSmH1}
\includegraphics[width=0.23\textwidth]{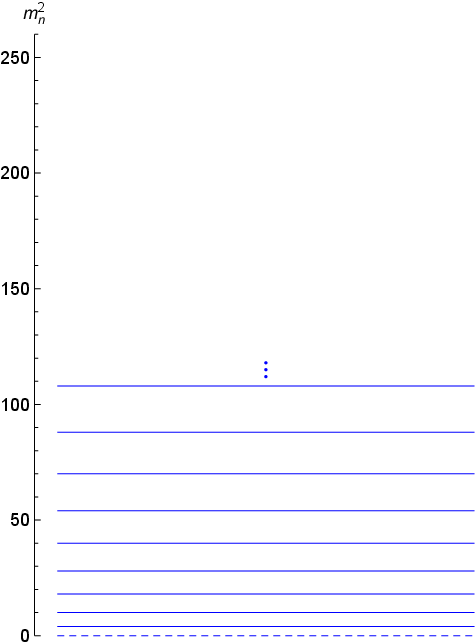}}
\subfigure[$H=1.5$]{\label{fig_AdSmH15}
\includegraphics[width=0.23\textwidth]{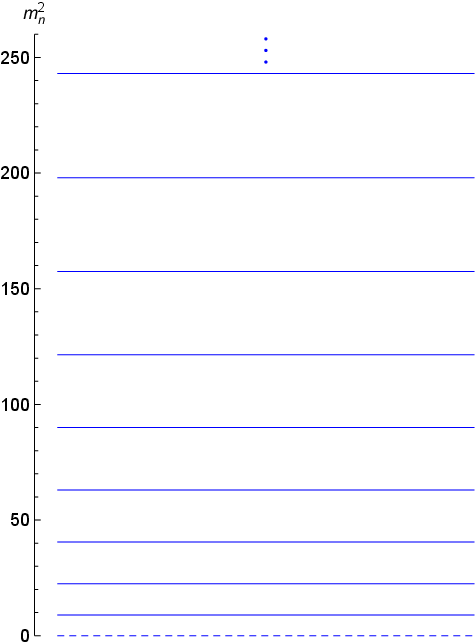}}
 \caption{ The $m_{n}^2$ spectrum of the KK modes for the
 $\mathrm{AdS_{4}}$ brane varies with different values of $H$.}\label{fig_AdSmH}
\end{center}\vskip -5mm
\end{figure}

The energy spectrum of bound states and the mass spectrum can be derived as follows:
\begin{eqnarray}
E_n = \left(\frac32 + n\right)^{2} H^2,\quad &&
m_{n}^{2} = n(n + 3)H^2.
\end{eqnarray}
We plot the $m_{n}^2$ spectrum of the KK modes in  Figure~\ref{fig_AdSmH}.  It is evident that the ground state corresponds to a massive mode with $m_1^{2} = 4H^{2}$.  All the KK modes are bound states and are localized on the thick AdS$_{4}$. The mass spectrum exhibits discontinuous.
Based on the preceding discussion, $m^2$ is always positive; further,  following the arguments presented in
Ref.~\cite{PRD_koyama}, it can be demonstrated that, in this context, all corresponding perturbation modes are stable.

\section{Perturbations and stability of the solutions  in the Einstein frame}\label{section4}

It is well-known that  a pure geometric $f(R)$ theory (known as the Jordan frame) is conformally equivalent to general relativity minimally coupled with a single canonical scalar field (referred to as the Einstein frame)~\cite{JS1988}. The linear perturbations of the latter have been extensively investigated in the literature~\cite{Giovannini2001,Giovannini2002_3,YY2016}. Therefore,  it is worthwhile to discuss the stability of our solutions in the Einstein frame. We introduce a conformal transformation defined as
\begin{equation}
\tilde{g}_{MN} = \Omega(z)^{2}g_{MN},
\end{equation}
where $\Omega(z)$ is a function of the coordinate $z$. We use a tilde to denote a quantity in the Einstein frame. Under this conformal transformation, the Ricci scalar is~\cite{Carroll2004}
\begin{equation}
\label{conformalR}
R = \Omega^2\tilde{R} + 8\tilde{g}^{MN}\Omega(\tilde{\nabla}_M\tilde{\nabla}_N\Omega)
 - 20 \tilde{g}^{MN}\tilde{\nabla}_M\Omega\tilde{\nabla}_N\Omega,
\end{equation}
where $\tilde{\nabla}_M$ denotes the covariant derivative associated with the conformal metric $\tilde{g}_{MN}$. We employ the relation $\sqrt{-g} = \Omega^{-5}\sqrt{-\tilde g}$ to rewrite the original gravitational action (\ref{action}) as
\begin{eqnarray}
\label{conformalS}
 S = \int {d^5}x\sqrt { -\tilde g} \left(\frac{{{}{f_R}}}{{2\kappa _5^2}}\Omega ^{ -5}R - {\Omega ^{ -5}}\sigma\right),
\end{eqnarray}
where $\sigma \equiv \frac{{{f_R}R - f(R)}}{{2\kappa _5^2}}$. By setting ${f_R} = {\Omega ^3}$, the original action (\ref{action}) can be simplified as~\cite{YY2016}
\begin{equation}
\label{actionSG}
S = \int {d^5}x\sqrt { - \tilde g} \left\{ \frac{1}{{2\kappa _5^2}}\tilde R - \frac{1}{2} \tilde{g}^{MN}\tilde{\nabla}_M\phi\tilde{\nabla}_N\phi - V(\phi)\right \}.
\end{equation}
Here, $\phi  = 2 \sqrt {\frac{{3}}{{\kappa _5^2}}} \ln \Omega$ and $V(\phi) = {\Omega^{-5}}\sigma$.
This action represents a minimally coupled scalar field in Einstein gravity. The linearization of the thick brane system with action (\ref{actionSG}) has been extensively examined in Ref.~\cite{Giovannini2001}, where the metric perturbations are classified into tensor, vector, and scalar modes. Each type of these modes evolves independently, and none of the perturbation equations relies explicitly on the form of $V(\phi)$.

To analyze fluctuations in the conformal metric, we express the metric  $\tilde{g}_{MN}$ in terms of the metric  $\hat{g}_{MN}$ and a conformal factor $\tilde{a}(z)$, as shown in the following equation
\begin{equation}
\tilde{g}_{MN} = \tilde{a}(z)^{2}\hat{g}_{MN},\label{gooe}
\end{equation}
where $\tilde{a}(z) \equiv e^{\tilde{A}(z)}$ and $\tilde{A}(z) = A(z) + \frac{1}{3}\ln f_{R}$.
Here, $e^{A(z)}$ represents the warp factor from the original metric in the Jordan frame. The conformal factor $\tilde{a}(z)$ corresponds to the new warp factor $e^{2\tilde{A}(z)}$ in the Einstein frame.

We further consider the metric fluctuations $\delta
g_{MN} = H_{MN} = \tilde{a}(z)^{2}h_{MN}$ from  Eq. (\ref{metric}). Under the
axial gauge $H_{5M} = 0$, the total metric can be expressed as
\begin{eqnarray}
 ds^{2} = \tilde{a}(z)^{2}\left[\big(\hat{g}_{\mu\nu}
    + h_{\mu\nu}\big)dx^\mu dx^\nu + dz^2\right],
\end{eqnarray}
in which $h_{\mu\nu} = h_{\mu\nu}(x^{\rho}, ~z)$ depends on all coordinates. After imposing the TT gauge condition $h^{\mu}_{\mu} = \hat{\nabla}^{\mu}h_{\mu\nu} = 0$, where $\hat{\nabla}_{\alpha}$ denotes the covariant derivative with
respect to $\hat{g}_{\mu\nu}$, the equation for the perturbation $h_{\mu\nu}$ takes the following form~\cite{PRD_koyama}
\begin{eqnarray}\label{dynamics_eq}
\left(\partial_{z}^{2} + 3 (\partial_{z}A)\partial_{z} +
 \hat{g}^{\alpha\beta}\hat{\nabla}_{\alpha}\hat{\nabla}_{\beta}
    - \frac{2}{3}\tilde{\lambda}_{4}\right)h_{\mu\nu} = 0.
\end{eqnarray}
Utilizing the KK decomposition $h_{\mu\nu}(x^{\rho}, z) = e^{-\frac{3}{2}\tilde{A}}\epsilon_{\mu\nu}(x^{\rho})\psi(z)$, where $\epsilon_{\mu\nu}(x^{\rho})$ satisfies the TT condition, we can derive from
Eq. (\ref{dynamics_eq})  the 4D equation $(\hat{g}^{\alpha\beta}\hat{\nabla}_{\alpha}\hat{\nabla}_{\beta} - \frac{2}{3}\tilde{\lambda}_{4})\epsilon_{\mu\nu}(x^{\rho}) = m^{2}\epsilon_{\mu\nu}(x^{\rho})$ and the Schr\"{o}dinger-like equation for the fifth-dimensional sector
\begin{eqnarray}\label{Schrodinger_eq}
\left[-\partial_{z}^{2} + V_\text{t}(z)\right]\psi(z) = m^{2}\psi(z),
\end{eqnarray}
where $m$ denotes the mass of the KK modes. The efffective potential is expressed as
\begin{align}\label{VQM}
V_{\text{t}}& = \frac{3}{2}\tilde{A}'' + \frac{9}{4}\tilde{A}'^{2}\nonumber\\
& = \frac{3}{2}A'' + \frac{9}{4}A'^{2} + \frac32A'\frac{f'_R}{f_R}
       - \frac14\frac{f_R'^2}{f_R^2} + \frac12\frac{f_R''}{f_R}.
\end{align}
Obviously, this outcome aligns with Eq. (\ref{Schrodingerpotential}).
The zero mode is expressed as
$\psi_{(0)}(z) \propto e^{\frac{3}{2}\tilde{A}(z)}$ \cite{PRD_koyama,RandallJHEP2001}, and must fulfill the normalization condition, given by
$\int |\psi_{(0)}|^{2} dz < \infty$.

\section{Conclusions}\label{section5}

In this paper, we investigate pure geometric branes embedded in a 5D spacetime within the framework of general $f(R)$ theory. The form of $f(R)$ is chosen as $f(R) = \sum_{i = 1}^{n} a_{i}\;R^{\,i} - 2 \Lambda_{5}$, without introducing background scalar fields. Based on the value of the 4D cosmological constant $\lambda_{4}$, we classify the branes into three cases:
 $\lambda_{4} = 0$ , $\lambda_{4} = 3 H^2$, and $\lambda_{4} = -3 H^2$, corresponding to $\mathcal{M}_{4}$, dS$_4$, and AdS$_4$ branes, respectively.
Assuming a  5D constant scalar curvature, we derive the corresponding brane solutions for each case. We then derive the perturbed Einstein equations and obtain the effective potential $W(z)$. The analysis of the associated Schr\"odinger-like equation reveals no gravitational modes with $m^2 < 0$, demonstrating the stability of these solutions against tensor perturbations.

For the $\mathcal{M}_{4}$ brane, the warp factor $e^{2A(z)}$ is discontinuous  and at most a $\delta$-function in second-order Einstein gravity, indicating the presence of a thin brane.  The effective potential involves a $\delta$-function,  leading to a single normalizable bound state mode, while the remaining eigenstates are continuum modes. These continuum modes asymptotically resemble plane waves, with their amplitudes suppressed near $z = 0$ due to the potential barrier.

For the dS$_4$ brane, the 5D scalar curvature $R$ is positive. Assuming $\alpha > 0$ , we find that the 5D spacetime is dS$_{5}$ when $0 \leqslant n \leqslant 4$, while it  is characterized as  AdS$_{5}$ spacetime for $n \geqslant 5$.
The effective potential is a P\"oschl-Teller potential,  with a minimum of $-\frac{3}{2}H^2$ at $z = 0$ and a maximum of $\frac{9}{4}H^2$ as $z = \pm\infty$, ensuring the presence of a mass gap. There are two bound states, corresponding to the zero mode and a massive mode. The continuum spectrum begins at $m^2 = \frac{9}{4} H^2$, with modes asymptotically behaving as plane waves representing delocalized KK gravitons.

In the case of the AdS$_4$ brane, the 5D scalar curvature $R$ is negative. Assuming $\alpha > 0$ , we find that the 5D spacetime is AdS$_{5}$ when $0 \leqslant n \leqslant 4$, while it is characterized as dS$_{5}$ for $n \geqslant 5$.
The effective potential is non-negative at the brane location, preventing the trapping of the zero mode while allowing massive modes to be trapped. The ground state is a massive mode with $m_1^{2} = 4H^{2}$. All massive modes are bound states localized on the thick AdS$_{4}$ brane, and the mass spectrum is discontinuous.

Finally, we examine tensor perturbations and the stability of these solutions in the Einstein frame. Through a conformal transformation, we reformulate the pure geometric $f(R)$ gravitational action into a general form. The results demonstrate that tensor fluctuations in the Einstein frame are consistent with those obtained directly in the Jordan frame.

\section*{Acknowledgments}

This work is supported by the National Natural
Science Foundation of China (Grants No. 11305119), the Natural Science Basic Research Plan in Shaanxi Province of China
(Program No. 2020JM-198), the Fundamental Research Funds for the Central Universities (Grants No. JB170502),
and the 111 Project (B17035).




\end{document}